\documentclass[10pt,final,journal,twocolumn,twoside,letterpaper]{IEEEtran}
\usepackage{cite}
\usepackage{graphicx}
\usepackage{amsmath}
\usepackage{times}
\usepackage{latexsym}
\usepackage{graphicx}
\usepackage{bm}
\usepackage{amssymb}
\usepackage[center]{caption2}
\usepackage{stfloats}
\usepackage{cases}
\usepackage{array}
\usepackage{setspace}
\usepackage{fancyhdr}

\newcounter{mytempeqncnt}
\allowdisplaybreaks[4]

\newcaptionstyle{mystyle1}{%
  \centering TABLE  \captiontext \par}
\captionstyle{mystyle1}
\newcaptionstyle{mystyle2}{%
  \captionlabel $.$ \: \captiontext \par}
\captionstyle{mystyle2}
\newcaptionstyle{mystyle3}{%
  \captionlabel $.$ \: \captiontext \par}
\captionstyle{mystyle3}

\renewcommand{\QED}{\QEDopen}

\begin{document}

\title{Simplified Frequency Offset Estimation for MIMO OFDM Systems}

\markboth{IEEE TRANSACTIONS ON VEHICULAR TECHNOLOGY, VOL. 57, NO. 5,
PP. 3246-3251, Sept. 2008}{JIANG \MakeLowercase{\text \it{et al.}}: Simplified
Frequency Offset Estimation for MIMO OFDM Systems}


\author{\authorblockN{Yanxiang~Jiang,~\IEEEmembership{Member,~IEEE,}
Hlaing~Minn,~\IEEEmembership{Senior Member,~IEEE,} Xiaohu~You,
and~Xiqi~Gao,~\IEEEmembership{Senior Member,~IEEE} }
\thanks{\small {Manuscript received June 19, 2007; revised October 24, 2007;
accepted December 14, 2007. The associate editor coordinating the
review of this manuscript and approving it for publication was Dr.
Jingxian Wu. The work of Yanxiang~Jiang, Xiaohu~You and Xiqi~Gao was
supported in part by National Natural Science Foundation of China
under Grants 60496310 and 60572072, the China High-Tech 863 Project
under Grant 2003AA123310 and 2006AA01Z264, and the International
Cooperation Project on Beyond 3G Mobile of China under Grant
2005DFA10360. The work of Hlaing~Minn was supported in part by the
Erik Jonsson School Research Excellence Initiative, the University
of Texas at Dallas, USA. This paper was presented in part at the
IEEE International Conference on Communications (ICC), Glasgow,
Scotland, June 2007.}}
\thanks{\small {Yanxiang~Jiang, Xiaohu~You and Xiqi~Gao are with the National
Mobile Communications Research Laboratory, Southeast University,
Nanjing 210096, China (e-mail: \{yxjiang, xhyu, xqgao\}
@seu.edu.cn).}}
\thanks{\small {Hlaing~Minn is with the Department of Electrical
Engineering, University of Texas at Dallas, TX 75083-0688, USA
(e-mail: hlaing.minn@utdallas.edu).}}
\thanks{Digital Object Identifier 00.0000/TVT.200X.00000.}}


\maketitle

\begin{abstract}
   This paper addresses a simplified frequency offset
estimator for multiple-input multiple-output (MIMO) orthogonal
frequency division multiplexing (OFDM) systems over frequency
selective fading channels. By exploiting the good correlation
property of the training sequences, which are constructed from the
Chu sequence, carrier frequency offset (CFO) estimation is obtained
through factor decomposition for the derivative of the cost function
with great complexity reduction. The mean-squared error (MSE) of the
CFO estimation is derived to optimize the key parameter of the
simplified estimator and also to evaluate the estimator performance.
Simulation results confirm the good performance of the
training-assisted CFO estimator.
\end{abstract}


\section{Introduction}

     Carrier frequency offset (CFO) estimation is an important issue for
both single-antenna and multiple-antenna orthogonal
frequency-division multiplexing (OFDM) systems \cite{Moose, David01,
FeifeiGao, Xiaoli, Simoens, Yxjiang3, YxjiangTWC}. Numerical
calculation of the maximum  likelihood (ML) CFO estimation is
computationally complicated since it requires a large point discrete
Fourier transform (DFT) operation and a time consuming line search.
Therefore, many papers have proposed reduced-complexity algorithms
\cite{David01, FeifeiGao, Simoens, Yxjiang3, YxjiangTWC}.
Especially, the search-free approaches were proposed in
\cite{FeifeiGao} \cite{Yxjiang3} \cite{YxjiangTWC}, where the
polynomial rooting is exploited to estimate the CFO. The solution
proposed in \cite{FeifeiGao} is based on computing the roots from
the derivative of the cost function whereas the solutions proposed
in \cite{Yxjiang3} \cite{YxjiangTWC} are based on computing the
roots directly from the cost function. However, both solutions still
need the complicated polynomial rooting operation, which is hard to
be implemented in practical OFDM systems \cite{Press}.

    In this paper, by further investigating the above search-free
approaches, a simplified CFO estimator is developed for MIMO OFDM
systems over frequency-selective fading channels. With the aid of
the training sequences generated from the Chu sequence \cite{Chu},
we propose to estimate the CFO via a simple polynomial factor. Thus,
the complicated polynomial rooting operation is avoided.
Correspondingly, the CFO estimator can be implemented via simple
additions and multiplications. To optimize the key parameter of the
simplified CFO estimator and also to evaluate the estimator
performance, the mean-squared error (MSE) of the CFO estimation is
derived.


    \textsl{Notations}: $( \cdot )_P $ denotes the remainder of the
number within the brackets modulo $P $. $\otimes$ and $\odot$ denote
the Kronecker product and Schur-Hadamard product, respectively.
$\Re(\cdot)$ and $\Im(\cdot)$ denote the real and imaginary parts of
the enclosed parameters, respectively. $\boldsymbol{x}^{(m)} $
denotes the $m$-cyclic-down-shift version of $\boldsymbol{x}$.
$\boldsymbol{F}_N$ and $\boldsymbol{I}_{N}$ denote the $N\times N$
unitary DFT matrix and identity matrix, respectively.
$\boldsymbol{e}_N^{k}$ denotes the $k$-th column vector of
$\boldsymbol{I}_N$. Unless otherwise stated, $0 \le \mu \le N_t-1$,
$0 \le \nu \le N_r-1$, $0 \le p \le P-1$ and $0 \le q \le Q-1$ are
assumed, where $Q=N/P$ with $(N)_{2P} = 0$.


\section{Signal Model}

    Consider a MIMO OFDM system with $N_t$ transmit antennas and
$N_r $ receive antennas and $N$ subcarriers. The training sequences
for CFO estimation are the same as in \cite{Yxjiang3}
\cite{YxjiangTWC}. Let $\boldsymbol{s}$ denote a length-$P$ Chu
sequence \cite{Chu}. Then, the $P \times 1$ pilot sequence vector at
the $\mu$-th transmit antenna is generated from $\boldsymbol{s}$ as
follows $ \tilde{\boldsymbol{s}}_{\mu} = \sqrt{Q/N_t}
\boldsymbol{F}_P \boldsymbol{s}^{(\mu M)} $, where $M = \lfloor
P/N_t \rfloor$. Define $\boldsymbol{\Theta}_{q} =
[\boldsymbol{e}_{N} ^{q}, \boldsymbol{e}_{N}^{q+Q}, \cdots,
\boldsymbol{e}_{N}^ {q+(P-1)Q}]$. Then, the $N \times 1$ training
sequence vector at the $\mu$-th transmit antenna is constructed as
follows $ \tilde{\boldsymbol{t}} _{\mu } = \boldsymbol{\Theta}_{
i_{\mu}} \tilde{\boldsymbol{s}}_{\mu}$, where $0 \leq i_{\mu} \leq
Q-1$ and $i_{\mu} = i_{\mu'} \ \mathrm{iff}. \ \mu = \mu '$. For
convenience, we refer to $\{ \tilde{\boldsymbol{t}}_{\mu } \}_{\mu
=0} ^{N_t-1}$ as the Chu sequence based training sequences (CBTS).

    Let $\boldsymbol{y}_{\nu}$ denote the $N
\times 1$ received vector at the $\nu$-th receive antenna after CP
removal. Let $\boldsymbol{h}^{(\nu, \mu)}$ denote the $L \times 1$
channel impulse response vector with $L$ being the maximum channel
length. Assume that $L$ is shorter than the length of cyclic prefix
(CP) $N_g$. Let $\tilde{\varepsilon}$ denote the frequency offset
normalized by the subcarrier frequency spacing. Define
\setlength{\arraycolsep}{0.0em}
\begin{eqnarray*}
\boldsymbol{y} &{}={}& [\boldsymbol{y}_{0}^T, \boldsymbol{y}_{1}^T,
\cdots, \boldsymbol{y}_{\nu}^T, \cdots, \boldsymbol{y}_{N_r-1}^T]^T,
\\
{\boldsymbol{h}} _{\nu} &{}={}& [ {({\boldsymbol{h}}^{(\nu,0)}) ^T},
\\
&{}{}& {({\boldsymbol{h}}^{(\nu,1)}) ^T}, \cdots,
{({\boldsymbol{h}}^{(\nu, \mu)}) ^T}, \cdots
{({\boldsymbol{h}}^{(\nu,N_t-1)}) ^T} ]^T,
\\
\boldsymbol{{h}} &{}={}& [\boldsymbol{{h}}_0^T,
\boldsymbol{{h}}_1^T, \cdots, \boldsymbol{{h}}_{\nu}^T, \cdots,
\boldsymbol{{h}}_{N_r-1}^T]^T,
\\
\boldsymbol{D}_{\bar{N}}(\tilde{\varepsilon}) &{}={}& \mathrm{diag}
\{[1,e^{j2\pi \tilde{\varepsilon} /N} , \cdots ,e^{j2\pi
\tilde{\varepsilon} (\bar{N}-1)/N}]^T \}.
\end{eqnarray*}\setlength{\arraycolsep}{5pt}\hspace*{-3pt}
Then, the cascaded received vector $\boldsymbol{y}$ over the $N_r$
receive antennas can be written as \cite{Yxjiang3} \cite{YxjiangTWC}
\begin{equation}\label{3-01}
\boldsymbol{y}  = \sqrt{N} e^{j{2\pi \tilde{\varepsilon} N_g } /
{N}} \{ \boldsymbol{I}_{N_r} \otimes [\boldsymbol{D}_N
(\tilde{\varepsilon} ) \boldsymbol{S}]\} \boldsymbol{h} +
\boldsymbol{w},
\end{equation}
where\setlength{\arraycolsep}{0.0em}
\begin{eqnarray*}
\boldsymbol{S} &{}={}& {\bar{\boldsymbol{F}}}^H \mathrm{diag} \{
[\tilde{\boldsymbol{s}}_ 0^T, \tilde{\boldsymbol{s}}_1^T, \cdots,
\tilde{\boldsymbol{s}} _{\mu}^T, \cdots, \tilde{\boldsymbol{s}}_
{N_t-1}^T] ^T\} \breve{\boldsymbol{F}},
\\
{\bar{\boldsymbol{F}}} &{}={}& [\boldsymbol{\Theta}_{i_0},
\boldsymbol{\Theta}_{i_1}, \cdots, \boldsymbol{\Theta} _{i_\mu},
\cdots, \boldsymbol{\Theta}_{i_{N_t-1}} ] ^T \boldsymbol{{F}}_N ,
\\
{{\boldsymbol{\breve{F}}}} &{}={}& [ \boldsymbol{e}_{N_t}^{0}
\otimes \boldsymbol{\Theta}_{i_0}^T, \boldsymbol{e}_{N_t}^{1}
\otimes \boldsymbol{\Theta}_{i_1}^T, \cdots,
\boldsymbol{e}_{N_t}^{\mu} \otimes \boldsymbol{\Theta}_{i_\mu}^T,
\\
&{}{}& \cdots, \boldsymbol{e}_{N_t}^{N_t -1} \otimes
\boldsymbol{\Theta}_{i_{N_t -1}}^T]  \{ \boldsymbol{I}_{N_t} \otimes
[\boldsymbol{F}_N [\boldsymbol{I}_L, \boldsymbol{0}_{L \times
(N-L)}]^T] \} .
\end{eqnarray*}\setlength{\arraycolsep}{5pt}\hspace*{-3pt}
and $\boldsymbol{w}$ is an $N_r N \times 1$ vector of uncorrelated
complex Gaussian noise samples with mean zero and equal variance of
$\sigma_w^2$.


\begin{figure*}[!b]
\hrulefill
\normalsize
\setcounter{mytempeqncnt}{\value{equation}}
\setcounter{equation}{18}
\begin{equation}\label{Phiqq}
\rho(\iota) = \left\{ {\begin{array}{*{20}l}
   {2\iota {\Re \{  ( \sum\nolimits_{\mu =0}
^{N_t-1} { z_{\mu}^{2\iota} } ) (\sum\nolimits_{\mu =0} ^{N_t-1} {
z_{\mu}^{-\iota} } )^2 \}  / {|\sum\nolimits_{\mu =0} ^{N_t-1}
 { z_{\mu}^{\iota} }|^2}}, } & {1 \le \iota \le Q/2 }  \\
{2 (Q - \iota) { \Re\{ (\sum\nolimits_{\mu =0} ^{N_t-1} {
z_{\mu}^{2\iota} }) ( \sum\nolimits_{\mu =0} ^{N_t-1} {
z_{\mu}^{\iota} } )^2   \} / { |\sum\nolimits_{\mu =0}
^{N_t-1} { z_{\mu}^{\iota} }|^2}},}    & {Q/2 < \iota \le Q - 1}  \\
\end{array}} \right.
\end{equation}
\setcounter{equation}{\value{mytempeqncnt}}
\end{figure*}

\section{Simplified CFO Estimator for MIMO OFDM Systems}

   By exploiting the periodicity property of CBTS, $\boldsymbol{y}$
can be stacked into the $Q \times N_r P$ matrix $\boldsymbol{Y} =
[\boldsymbol{Y}_0, \boldsymbol{Y}_1, \cdots, \boldsymbol{Y}_\nu,
\cdots, \boldsymbol{Y}_{N_r-1}]$ with its element given by
$[\boldsymbol{Y}_{\nu}] _{q,p} = [((\boldsymbol{e} _{N_r} ^{\nu}) ^T
\otimes \boldsymbol{I}_N) \boldsymbol{y}]_{qP+p}$. Define
\setlength{\arraycolsep}{0.0em}
\begin{eqnarray*}
\boldsymbol{b}_{\mu} &{}={}& [1, e ^{j {2\pi
(\tilde{\varepsilon}+i_\mu)}/{Q} }, \\
&{}{}& \cdots, e ^{j {2\pi (\tilde{\varepsilon}+i_\mu)} q /{Q} },
\cdots, e ^{j {2\pi (\tilde{\varepsilon}+i_\mu)(Q-1)}/{Q} }]^T,
\\
\boldsymbol{B}(\tilde{\varepsilon}) &{}={}& [\boldsymbol{b}_0,
\boldsymbol{b}_1, \cdots, \boldsymbol{b}_{\mu}, \cdots,
\boldsymbol{b}_{N_t -1}].
\end{eqnarray*}\setlength{\arraycolsep}{5pt}\hspace*{-3pt}
Then, $\boldsymbol{Y}$ can be expressed in the following equivalent
form \cite{Yxjiang3} \cite{YxjiangTWC}
\begin{equation}
\boldsymbol{Y} = \boldsymbol{B}(\tilde{\varepsilon})\boldsymbol{X} +
\boldsymbol{W},
\end{equation}
where 
\setlength{\arraycolsep}{0.0em}
\begin{eqnarray*}
\boldsymbol{X} &{}={}& [\boldsymbol{X}_{0}, \boldsymbol{X}_{1},
\cdots, \boldsymbol{X}_{\nu}, \cdots, \boldsymbol{X}_{N_r -1}],
\\
\boldsymbol{X}_{\nu} &{}={}& [\boldsymbol{x}^{(\nu,0)},
\boldsymbol{x}^{(\nu,1)}, \cdots, \boldsymbol{x}^{(\nu, \mu)},
\cdots, \boldsymbol{x}^{(\nu, N_t-1)}]^T,
\\
\boldsymbol{x}^{(\nu, \mu)} &{}={}& {\sqrt{P}} e^{j {2\pi
\tilde{\varepsilon} N_g }/{N}} \boldsymbol{D}_{P}
(\tilde{\varepsilon}+i_\mu ) \\
&{}{}& \times \boldsymbol{F}_P^H
\mathrm{diag}\{\tilde{\boldsymbol{s}}_{\mu}\} \boldsymbol{\Theta}
_{i_\mu}^T \boldsymbol{F}_N [\boldsymbol{I}_L, \boldsymbol{0}_{L
\times (N-L)}]^T \boldsymbol{h}^{(\nu,\mu)},
\end{eqnarray*}\setlength{\arraycolsep}{5pt}\hspace*{-3pt}
and $\boldsymbol{W}$ is the $Q \times N_r P$ matrix generated from
$\boldsymbol{w}$ in the same way as $\boldsymbol{Y}$.

    According to the multivariate statistical theory, the log-likelihood
function of $\boldsymbol{Y}$ conditioned on
$\boldsymbol{B}(\varepsilon)$ and $\boldsymbol{X}$ with
$\varepsilon$ denoting a candidate CFO can be obtained as follows
\begin{equation}
\ln p(\boldsymbol{Y} | \boldsymbol{B}(\varepsilon), \boldsymbol{X})
= -\sigma _w ^{-2}\mathrm{Tr}\{[\boldsymbol{Y}-
\boldsymbol{B}(\varepsilon) \boldsymbol{X}] [\boldsymbol{Y}-
\boldsymbol{B}(\varepsilon) \boldsymbol{X}]^H\}.
\end{equation}
Exploit the condition $i_{\mu} = i_{\mu '} \ \mathrm{iff}. \ \mu =
\mu ' $. Then, after some straightforward manipulations, we can
obtain the reformulated log-likelihood function conditioned on
$\varepsilon$ as follows
\begin{equation}\label{rlikelihood}
\ln p(\boldsymbol{Y} | \varepsilon) =
\mathrm{Tr}[\boldsymbol{B}^H(\varepsilon)
\hat{\boldsymbol{R}}_{\boldsymbol{Y}\boldsymbol{Y}}
\boldsymbol{B}(\varepsilon)],
\end{equation}
where $\hat{\boldsymbol{R}}_{\boldsymbol{Y}\boldsymbol{Y}} =
\boldsymbol{Y} \boldsymbol{Y}^H$. Direct grid searching from
(\ref{rlikelihood}) yields the ML estimate, however, this approach
is computationally quite expensive. In order to compute the CFO
efficiently, we will propose a simplified CFO estimator for MIMO
OFDM systems subsequently.


    Define
$z = e^{j2\pi \varepsilon /Q}$, $z_{\mu} = e^{j2\pi i_{\mu} /Q}$,
$\boldsymbol{b}(z) = [1, z, \cdots, z^q, \cdots, z^{Q-1}]^T$. Then,
by exploiting the Hermitian property of $\hat{\boldsymbol{R}}
_{\boldsymbol{Y}\boldsymbol{Y}}$, the log-likelihood function in
(\ref{rlikelihood}) can be transformed into the following equivalent
form
\begin{multline}\label{fz}
f(z) = \boldsymbol{c}^T \left\{ \left[\sum\limits _{\mu=0} ^{N_t-1}
{\boldsymbol{b} (z_{\mu})} \right] \odot \boldsymbol{b}(z) \right\}
\\ +  \boldsymbol{c}^H \left\{ \left[\sum\limits _{\mu=0} ^{N_t-1}
{\boldsymbol{b} (z_{\mu}^{-1})} \right] \odot \boldsymbol{b}(z^{-1})
\right\},
\end{multline}
where $\boldsymbol{c}$ is a $Q \times 1$ vector with its $q$-th
element given by $[\boldsymbol{c}]_q = \sum\nolimits_{j-i = q}
{[\hat{\boldsymbol{R}}_{\boldsymbol{Y}\boldsymbol{Y}}]_{i,j}}$. It
can be seen from its definition that the $q$-th element of
$\boldsymbol{c}$ corresponds to the summation of the $q$-th upper
diagonal elements of $\hat{\boldsymbol{R}} _{\boldsymbol{Y}
\boldsymbol{Y}}$. Taking the first-order derivative of $f(z)$ with
respect to $z$ yields
\begin{multline}\label{equ6}
f'(z) = z^{-1} \left\{ \boldsymbol{c}^T \left\{ \left[\sum\limits
_{\mu=0} ^{N_t-1} {\boldsymbol{b} (z_{\mu})} \right] \odot
\boldsymbol{b}(z) \odot \boldsymbol{q} \right\} \right. \\
\left. - \boldsymbol{c}^H \left\{ \left[\sum\limits _{\mu=0}
^{N_t-1} {\boldsymbol{b} (z_{\mu}^{-1})} \right] \odot
\boldsymbol{b}(z^{-1}) \odot \boldsymbol{q} \right\} \right\},
\end{multline}
where $\boldsymbol{q} = [0, 1, \cdots, q, \cdots, Q-1] ^T$. By
letting the derivative of the log-likelihood function $f'(z)$ be
zero, the solutions for all local minima or maxima can be obtained.
Put these solutions back into the original log-likelihood function
$f(z)$ and select the maximum by comparing all the solutions
obtained in the previous stage. The improved blind CFO estimator
exploiting the above mathematical rule has been addressed for
single-antenna OFDM systems in \cite{FeifeiGao}. Although the
search-free approach has a relatively lower complexity, it still
requires a complicated polynomial rooting operation, which is hard
to be implemented in practical OFDM systems. With the aid of the
CBTS training sequences, we will show in the following that the
polynomial rooting operation can be avoided for the training aided
CFO estimation in MIMO OFDM systems.

    Assume that $P \geq L$, the channel taps remain constant during
the training period, and the channel energy is mainly concentrated
in the first $M$ taps with $M<L$. Then, we have (see Appendix I for
the details)
\begin{multline}\label{ghz}
\boldsymbol{c}^H \left\{ \left[\sum\limits _{\mu=0} ^{N_t-1}
{\boldsymbol{b} (z_{\mu}^{-1})} \right] \odot \boldsymbol{b}(z^{-1})
\odot \boldsymbol{q} \right\} \\
= z^{-Q} \kappa(\iota) \cdot
\boldsymbol{c}^T \left\{ \left[\sum\limits _{\mu=0} ^{N_t-1}
{\boldsymbol{b} (z_{\mu})} \right] \odot \boldsymbol{b}(z) \odot
\boldsymbol{q} \right\} ,
\end{multline}
where $\kappa(\iota) = {\iota [\boldsymbol{c}]^*_{\iota}}
/{[(Q-\iota) [\boldsymbol{c}]_{Q-\iota}]}$ with $1 \leq \iota \le
Q-1$, and the parameter $\iota$ denotes the index of the upper
diagonal of $\hat{\boldsymbol{R}} _{\boldsymbol{Y} \boldsymbol{Y}}$.
From (\ref{ghz}), it follows immediately that $f'(z)$ can be
decomposed as follows
\begin{equation}\label{equ8}
f'(z) = z^{-(Q+1)} [z^Q - \kappa(\iota)] \cdot \boldsymbol{c}^T
\left\{ \left[\sum\limits _{\mu=0} ^{N_t-1} {\boldsymbol{b}
(z_{\mu})} \right] \odot \boldsymbol{b}(z) \odot \boldsymbol{q}
\right\}.
\end{equation}
Define $\tilde{z} = e^{j2\pi \tilde{\varepsilon} /Q} $. Assume $N_t
< Q$. Then, with (\ref{coeff}) as shown in Appendix I, we have (see
Appendix II for the details)
\begin{equation}\label{gzl0}
\boldsymbol{c}^T \left\{ \left[\sum\limits _{\mu=0} ^{N_t-1}
{\boldsymbol{b} (z_{\mu})} \right] \odot \boldsymbol{b}(\tilde{z})
\odot \boldsymbol{q} \right\}  > 0, \ f'(\tilde{z}) = 0.
\end{equation}
It follows from (\ref{equ8}) and (\ref{gzl0}) that $z = \tilde{z}$
is one of the roots of both $f'(\tilde{z}) = 0$ and $z^Q -
\kappa(\iota) = 0$. Unlike $f'(\tilde{z}) = 0$, the roots of $z^Q -
\kappa(\iota) = 0$ can be calculated without the polynomial rooting
operation. Therefore, by solving the simple polynomial equation $z^Q
- \kappa(\iota) = 0$, the CFO estimate can be obtained efficiently
as follows
\begin{equation}
\hat{\varepsilon} = \mathop {\arg \max } \limits_{\varepsilon \in
\{\varepsilon_{q} \} _{q=0}^{Q-1} } \{ f(z) | z = e^{j2\pi
\varepsilon /Q} \},
\end{equation}
where $\varepsilon_{q} = \mathrm{arg} \{\kappa(\iota)\} / {(2 \pi)}
+ q - {Q}/{2}$. It can be calculated that the main computational
complexity of the simplified CFO estimator is $4N_rN Q + 8 Q^2$.
Compared with the CFO estimator in [6] [7], whose main computational
complexity is $4N_rN \log_2 N + 9 Q^3 + 64/3 (Q-1)^3$, the
complexity of the simplified CFO estimator is generally lower.
Furthermore, since the polynomial rooting operation is avoided, the
simplified CFO estimator can be implemented via simple additions and
multiplications, which is more suitable for practical OFDM systems.
Note that $\iota$ is a key parameter for the proposed CFO estimator.
We will show in the following how to determine the optimal $\iota$.

\section{Performance Analysis and Parameter Optimization}
    To optimize $\iota$ and also to evaluate the estimation
accuracy, we first derive the MSE of the simplified CFO estimator.
Invoking the definition of $\hat{\boldsymbol{R}} _{\boldsymbol{Y}
\boldsymbol{Y}}$, we can readily obtain
\begin{equation}
\hat{\boldsymbol{R}}_{\boldsymbol{Y}\boldsymbol{Y}} \doteq N_r P
\sigma _x^2 \boldsymbol{B} (\tilde{\varepsilon}) \boldsymbol{B}^H
(\tilde{\varepsilon}) +
\hat{\boldsymbol{R}}_{\boldsymbol{Y}\boldsymbol{W}} +
\hat{\boldsymbol{R}}_{\boldsymbol{W}\boldsymbol{W}},
\end{equation}
where $\sigma _x^2 = \mathrm{E} [ |[\boldsymbol{x}^{(\nu,
\mu)}]_p|^2 ]$, $\hat{\boldsymbol{R}}_{\boldsymbol{Y}\boldsymbol{W}}
= \boldsymbol{B} (\tilde{\varepsilon}) \boldsymbol{X}
\boldsymbol{W}^H + \boldsymbol{W} \boldsymbol{X}^H \boldsymbol{B}^H
(\tilde{\varepsilon})$, $\hat{\boldsymbol{R}}_{\boldsymbol{W}
\boldsymbol{W}} = \boldsymbol{W} \boldsymbol{W} ^H$. Assume
\begin{multline*}
\mathrm{E} \{ [\boldsymbol{h}^{(\nu,\mu)}]_l^*
[\boldsymbol{W}]_{i,j} \}= 0, \ \mathrm{E} \{
[\boldsymbol{h}^{(\nu,\mu)}]_l^* [\boldsymbol{h}^{(\nu ',\mu
')}]_{l'} \} = 0, \\
\forall (\nu,\mu) \ne (\nu ',\mu ').
\end{multline*}
Then, for $\ i \neq j$ and $\ i' \neq j'$, it can be concluded
directly from their definitions that
\begin{equation}
\mathrm{E} [ [\hat{\boldsymbol{R}} _{\boldsymbol{Y}
\boldsymbol{W}}]_{i,j} ] = \mathrm{E} [ [\hat{\boldsymbol{R}}
_{\boldsymbol{W} \boldsymbol{W}}]_{i,j} ] = \mathrm{E} [
[\hat{\boldsymbol{R}}_{\boldsymbol{Y}\boldsymbol{W}}]_{i,j}^*
[\hat{\boldsymbol{R}}_{\boldsymbol{W}\boldsymbol{W}}]_{i',j'} ] = 0,
\end{equation}
\begin{equation}
\mathrm{E} [ |[\hat{\boldsymbol{R}}_{\boldsymbol{Y}
\boldsymbol{W}}]_{i,j}|^2 ] = 2N_tN_r P \sigma _x^2 \sigma _w^2, \
\mathrm{E} [ |[\hat{\boldsymbol{R}}_{\boldsymbol{W}
\boldsymbol{W}}]_{i,j}|^2 ] = N_r P \sigma _w^4.
\end{equation}
Invoking the definition of $\boldsymbol{c}$, we have
\begin{equation}
[\boldsymbol{c}]_\iota = N_r P \sigma _x^2 (Q- \iota)
\tilde{z}^{-\iota} \sum\limits_{\mu =0} ^{N_t-1} { z_{\mu}^{-\iota}
} + \alpha_\iota + \beta_\iota,
\end{equation}
where $\alpha_\iota = \sum\nolimits_{j-i = \iota}
{[\hat{\boldsymbol{R}}_{\boldsymbol{Y} \boldsymbol{W}}]_{i,j}}$,
$\beta_\iota = \sum\nolimits_{j-i = \iota}
{[\hat{\boldsymbol{R}}_{\boldsymbol{W} \boldsymbol{W}}]_{i,j}}$. It
follows immediately from its definition that $\kappa(\iota)$ can be
expressed as
\begin{equation}\label{k-27}
\kappa(\iota) = \frac{ \tilde{z}^{\iota} + \zeta _\iota }{
\tilde{z}^{-Q+\iota} + \eta_\iota } = \frac{ \tilde{z}^Q +
\xi_\iota}{|\tilde{z}^{-Q+\iota} + \eta_\iota|^2 },
\end{equation}
where \setlength{\arraycolsep}{0.0em}
\begin{eqnarray*}
\zeta _\iota &{}={}& \{ \alpha_\iota^* + \beta_\iota^* \} / { [N_r P
(Q-\iota) \sigma _x^2 \sum\nolimits_{\mu =0} ^{N_t-1} {
z_{\mu}^{\iota} } ]},
\\
\eta_\iota &{}={}& { \{ \alpha_{Q-\iota} + \beta_{Q-\iota} \} } / {
[ N_r P  \iota \sigma _x^2 \sum\nolimits_{\mu =0} ^{N_t-1} {
z_{\mu}^{\iota} } ] }, \\
\xi_\iota &{}= {}& \tilde{z}^{Q-\iota} \zeta _\iota +
\tilde{z}^{\iota} \eta_\iota^* + \zeta _\iota \eta_\iota^*.
\end{eqnarray*}\setlength{\arraycolsep}{5pt}\hspace*{-3pt}
\vspace*{-10pt}

    From (\ref{k-27}), we can see that the MSE of the estimated CFO
is highly related to the variances of $\zeta _\iota$, $\eta_\iota$
and $\xi_\iota$. By invoking their definitions, the variances of
$\zeta _\iota$ and $\eta_\iota$ can be directly calculated as
follows
\begin{equation}
\mathrm{var}\{ \zeta _\iota \} = \frac{ 2 N_t \gamma^{-1} +
\gamma^{-2} } { N_r P (Q-\iota) \left|\sum\limits_{\mu =0} ^{N_t-1}
{ z_{\mu}^{\iota} } \right|^2 }, \ \mathrm{var}\{ \eta _\iota \} =
\frac{ 2 N_t \gamma^{-1} + \gamma^{-2} }{ N_r P \iota
\left|\sum\limits_{\mu =0} ^{N_t-1} { z_{\mu}^{\iota} } \right|^2 },
\end{equation}
where $\gamma = \sigma _x^2/ \sigma _w^2$. When $\gamma \gg 1$,
i.e., the signal-to-noise ratio (SNR) is large enough, we have
\begin{equation}
\mathrm{var}\{ \zeta _\iota \} \gg \mathrm{var}\{ \zeta _\iota \eta
_\iota^* \}, \ \mathrm{var}\{ \eta _\iota \} \gg \mathrm{var}\{
\zeta _\iota \eta _\iota^* \}.
\end{equation}
Accordingly, the variance of $ \xi _\iota $ can be approximated as
follows \setlength{\arraycolsep}{0.0em}
\begin{eqnarray}
\mathrm{var}\{ \xi _\iota \} \ &\doteq& \ \mathrm{var}\{ \zeta
_\iota \} + \mathrm{var}\{ \eta_\iota \} + \mathrm{E} [
\tilde{z}^{Q-2\iota} \zeta _\iota \eta_\iota ] + \mathrm{E} [
\tilde{z}^{-Q+2\iota} \zeta _\iota^*
\eta_\iota^* ], \nonumber \\
\ & = &\ \frac{ 2 [N_t Q +\rho(\iota)] \gamma^{-1} + Q \gamma^{-2}
}{ N_r P \iota (Q-\iota) \biggl|\sum\limits_{\mu =0} ^{N_t-1} {
z_{\mu}^{\iota} } \biggl|^2 },
\end{eqnarray} \setlength{\arraycolsep}{5pt}
where $\rho(\iota)$ is shown in (19) at the bottom of the page.
\stepcounter{equation} Note that $\rho(\iota)$ is a nonlinear
function with respect to $\iota$ and $z_{\mu}$. When $\mathrm{var}\{
\xi _\iota \} \ll 1$, which is a reasonable assumption for the
practical systems, it follows immediately from (\ref{k-27}) that
\setlength{\arraycolsep}{0.0em}
\begin{eqnarray}
\hat \varepsilon \ & = & \ \tilde \varepsilon + \frac{1} {2\pi}
\mathrm{arg}\{ 1 + e^ {-j 2 \pi \tilde{\varepsilon}} \xi _\iota \}
\nonumber \\
\ & \doteq & \ \tilde \varepsilon + \frac{1}{2\pi} \Im \{ e^ {-j 2
\pi \tilde{\varepsilon}} \xi _\iota \}.
\end{eqnarray} \setlength{\arraycolsep}{5pt}
Then, the MSE of the estimated CFO can be readily obtained as
follows
\begin{equation}\label{var_epsilon}
\mathrm{MSE}\{ \hat {\varepsilon} \} \doteq \frac{1}{8 \pi ^2}
\mathrm{var}\{ \xi _\iota \} \doteq \frac{ 2 [N_t Q +\rho(\iota)]
\gamma^{-1} + Q \gamma^{-2} }{ 8 \pi ^2 N_r P \iota (Q-\iota)
\left|\sum\limits_{\mu =0} ^{N_t-1} { z_{\mu}^{\iota} } \right|^2 }.
\end{equation}
It can be seen from (\ref{var_epsilon}) that $\mathrm{MSE}\{ \hat
{\varepsilon} \}$ depends on $\iota$ for fixed $\{ i_{\mu} \}_{\mu
=0 }^{N_t-1}$, $N_t, N_r, P, Q$ and $\gamma$. To obtain a better
estimator performance, we can optimize the parameter $\iota$ based
on (\ref{var_epsilon}).

\section{Simulation Results}

\begin{figure}[!b]
\centering 
\includegraphics[width=0.45\textwidth]{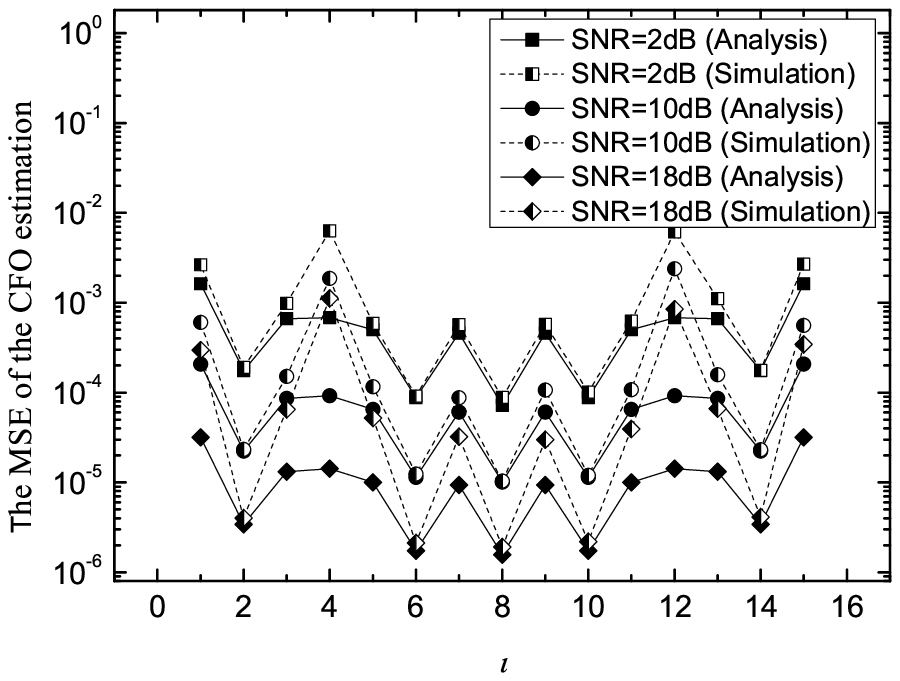}
\caption{The MSE of the proposed CFO estimator as a function of
$\iota$ with $\{ i_{\mu} \}_ {\mu =0}^ {N_t-1} = \{ 3, 5, 11\}$.}
\label{fig1}
\end{figure}

\begin{figure}[!b]
\centering 
\includegraphics[width=0.45 \textwidth]{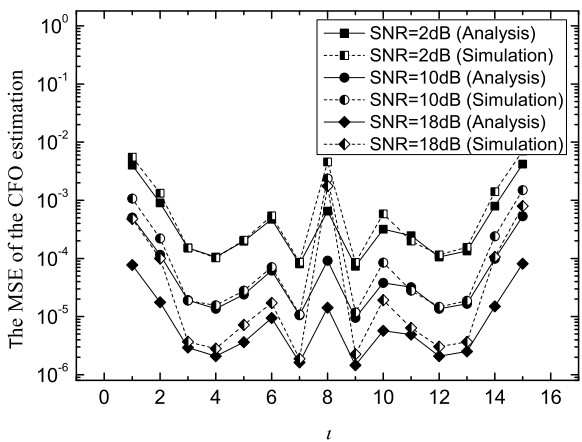}
\caption{The MSE of the proposed CFO estimator as a function of
$\iota$ with $\{ i_{\mu} \}_ {\mu =0}^ {N_t-1} = \{ 3, 7, 14\}$.}
\label{fig2}
\end{figure}

    Numerical results are provided to verify the analytical results
and also to evaluate the performance of the proposed CFO estimator.
The considered MIMO OFDM system is of bandwidth $20$MHz and carrier
frequency $5$GHz with $N=1024$ and $N_g = 80$. Each of the channels
is with $6$ independent Rayleigh fading taps, whose relative
average-powers and propagation delays are $\{0, -0.9, -4.9, -8.0,
-7.8, -23.9\}$dB and $\{0, 4, 16, 24, 46, 74\}$ samples,
respectively. The other parameters are as follows: $P=64$, $Q=16$,
$N_t =3$, $N_r =2$, $\tilde{\varepsilon} \in (-Q/2,Q/2)$.

    Figs. \ref{fig1} and \ref{fig2} present the MSE
of the proposed CFO estimator as a function of $\iota$ with $\{
i_{\mu} \}_ {\mu =0}^ {N_t-1} = \{ 3, 5, 11\}$ and $\{ 3, 7, 14\}$,
respectively. The solid and dotted curves are the results from
analysis and Monte Carlo simulations, respectively. It can be
observed that the results from analysis agree quite well with those
from simulations except when the actual MSE of the estimate is very
large. It can also be observed that $\mathrm{MSE}\{ \hat \varepsilon
\}$ achieves its minimum for $\iota = 6, 8 ,10$ with $\{ i_{\mu} \}_
{\mu =0}^ {N_t-1} = \{ 3, 5, 11\}$ and for $\iota = 7, 9$ with $\{
i_{\mu} \}_ {\mu =0}^ {N_t-1} = \{ 3, 7, 14\}$. These observations
imply that we can obtain the optimum value of the parameter $\iota$
from the analytical results after $\{ i_{\mu} \}_ {\mu =0}^ {N_t-1}$
is determined.

\begin{figure}[!b]
\centering 
\includegraphics[width=0.45\textwidth]{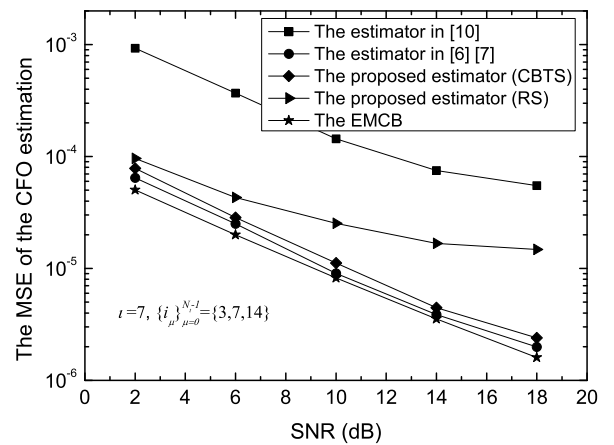}
\caption{The MSE of the different CFO estimators as a function of
SNR.} \label{fig3}
\end{figure}

    Depicted in Fig. \ref{fig3} is the performance comparison
between the proposed CFO estimator ($\iota=7$, $\{ i_{\mu} \}_ {\mu
=0}^ {N_t-1} = \{ 3, 7, 14\}$) and the estimator in \cite{Yxjiang3}
\cite{YxjiangTWC} and \cite{Stuber}. In order to substantiate that
the training sequences generated from the Chu sequence do help to
improve the estimation accuracy, the performance of the proposed
estimator with random sequences (RS), whose elements are generated
randomly, is included. Also included as a performance benchmark is
the extended Miller and Chang bound (EMCB) \cite{Gini}
\cite{Minn_Periodic}, which is obtained by averaging the snapshot
Cramer-Rao bound (CRB) over independent channel realizations as
follows
\begin{multline}
\mathrm{EMCB}_\varepsilon  \\
= \mathrm{E} \left \{
\frac{N\sigma_w^2} {8 \pi^2 \boldsymbol{h}^H
\boldsymbol{\mathcal{X}}^H \boldsymbol{\mathcal{B}}
[\boldsymbol{I}_{N_rN} - \boldsymbol{\mathcal{X}}
(\boldsymbol{\mathcal{X}}^H \boldsymbol{\mathcal{X}})^{-1}
\boldsymbol{\mathcal{X}}^H] \boldsymbol{\mathcal{B}}
\boldsymbol{\mathcal{X}} \boldsymbol{h} } \right\},
\end{multline}
where $\boldsymbol{\mathcal{X}} = \boldsymbol{I}_{N_r} \otimes
\boldsymbol{S}$, $\boldsymbol{\mathcal{B}} = \boldsymbol{I}_{N_r}
\otimes \mathrm{diag}\{[N_g, N_g+1, \cdots, N_g +N -1]^T \}$. We
resort to Monte Carlo simulation for its evaluation. It can be
observed that the performance of the proposed estimator with CBTS is
far better than that in \cite{Stuber} and slightly worse than that
in \cite{Yxjiang3} \cite{YxjiangTWC}, and its performance also
approaches the EMCB which verifies its high estimation accuracy. It
can also be observed that the performance of the proposed CFO
estimator with CBTS is far better than that with RS, which should be
attributed to the good correlation property of CBTS.

\section{Conclusions}

    In this paper, we have presented a low complexity CFO estimator
for MIMO OFDM systems with the training sequences generated from the
Chu sequence. The MSE of the CFO estimation has been developed to
evaluate the estimator performance and also to optimize the key
parameter. By exploiting the optimized parameter from the estimation
MSE, our CFO estimator with CBTS yields good performance.

\appendices
\section{} 

    This appendix presents the proof of (\ref{ghz}). It follows
immediately that the polynomials on the two sides of (\ref{ghz}) are
both $(Q-1)$-degree. Therefore, in order to validate the
relationship (\ref{ghz}), we only need to prove that the
corresponding polynomial coefficients are pairwise equal.

    Ignoring the noise items, $\hat{\boldsymbol{R}} _{\boldsymbol{Y}
\boldsymbol{Y}}$ can be expressed as follows
\begin{equation}
\hat{\boldsymbol{R}}_{\boldsymbol{Y} \boldsymbol{Y}} =
\boldsymbol{B} (\tilde{\varepsilon}) \hat{\boldsymbol{R}}_
{\boldsymbol{X} \boldsymbol{X}} \boldsymbol{B}^H
(\tilde{\varepsilon}),
\end{equation}
where $\hat{\boldsymbol{R}}_{\boldsymbol{X}\boldsymbol{X}} =
\boldsymbol{X} \boldsymbol{X}^H$. Define
\[
\varpi_{\mu, \mu '} = { (i_{\mu }-i_{\mu '})}/{Q}, \
\boldsymbol{s}_{\mu} = \sqrt{N_t/Q} \boldsymbol{F}_P^H
\tilde{\boldsymbol{s}}_{\mu}.
\]
Then, with the assumptions that $P \ge L$ and the channel taps
remain constant during the training period, we can readily obtain
\begin{equation}\label{Rxxmumu}
[\hat{\boldsymbol{R}}_{\boldsymbol{X}\boldsymbol{X}}]_{\mu, \mu '} =
\frac{N}{N_t } \sum\limits_{\nu =0}^{N_r -1}\sum\limits_{l=0} ^{L-1}
\sum\limits_{l' =0}^{L-1} \left\{[\boldsymbol{h}^{(\nu, \mu)}]_l
[\boldsymbol{h} ^{(\nu, \mu ')}]_{l'}^* [\boldsymbol{A}^{(\mu, \mu
')}]_{l, l'} \right\},
\end{equation}
where $[\boldsymbol{A}^{(\mu, \mu ')}]_{l, l'} = (\boldsymbol{s}_
{\mu}^{(l)})^T \boldsymbol{D}_P(\varpi_{\mu, \mu '} Q)
(\boldsymbol{s}_{\mu '} ^{(l')})^*$. With the aim of complexity
reduction, $[\hat{\boldsymbol{R}}_{\boldsymbol{X}
\boldsymbol{X}}]_{\mu, \mu '}$ is replaced with its expected value.
Exploiting the good correlation property of CBTS, which is inherited
from the Chu sequence, we obtain
\begin{equation}
[\hat{\boldsymbol{R}}_{\boldsymbol{X}\boldsymbol{X}}]_{0,0} =
[\hat{\boldsymbol{R}}_{\boldsymbol{X}\boldsymbol{X}}]_{1,1} = \cdots
= [\hat{\boldsymbol{R}}_ {\boldsymbol{X}\boldsymbol{X}}] _{N_t-1,
N_t -1}.
\end{equation}
Let $[\boldsymbol{s}]_p = e^{j\pi \upsilon p^2 / P}$ with $\upsilon$
being coprime with $P$. Define $p_{\mu, l} = \mu M + l$. Then, we
have \setlength{\arraycolsep}{0.0em}
\begin{eqnarray*}
[\boldsymbol{A}^{(\mu, \mu ')}]_{l, l'} &{}={}&
(-1)^{\upsilon(p_{\mu, l} - p_{\mu', l'}) +1} e^{j \pi \upsilon
(p_{\mu, l} ^2 - p_{\mu', l'}^2) / P}
\\
&{}{}& \times e^{-j \pi (P-1)[\upsilon(p_{\mu, l} - p_{\mu', l'}) -
\varpi_{\mu, \mu '}]/P} \sin (\pi \varpi_{\mu, \mu '}) \\
&{}{}& / \sin \{ \pi [\upsilon(p_{\mu, l} - p_{\mu', l'}) -
\varpi_{\mu, \mu '}]/P \}.
\end{eqnarray*}\setlength{\arraycolsep}{5pt}\hspace*{-3pt}

It follows immediately that
\begin{equation}
\bigl|[\boldsymbol{A}^ {(\mu, \mu ')}] _{l, l'} \bigl|_{p_{\mu, l} -
p_{\mu', l'} \ne 0} \ll \bigl| [\boldsymbol{A}^{(\mu, \mu ')}]_{l,
l'} \bigl|_{p_{\mu, l} - p_{\mu', l'} = 0}.
\end{equation}
With the assumption that the channel energy is mainly concentrated
in the first $M$ taps,
$\bigl|[\hat{\boldsymbol{R}}_{\boldsymbol{X}\boldsymbol{X}}]_{\mu,
\mu '}\bigl|_{\mu \ne \mu '}$ can be made very small with CBTS,
which yields
\begin{equation}
\bigl|[\hat{\boldsymbol{R}}_ {\boldsymbol{X} \boldsymbol{X}}]_ {\mu,
\mu '}\bigl|_{\mu \ne \mu '} \ll [\hat{\boldsymbol{R}}
_{\boldsymbol{X} \boldsymbol{X}}]_{\mu, \mu}.
\end{equation}
Then, we have $\hat{\boldsymbol{R}}_{\boldsymbol{X}\boldsymbol{X}}
\doteq N_r P \sigma _x^2 \boldsymbol{I}_{N_t}$, where $\sigma _x^2 =
\mathrm{E} [ |[\boldsymbol{x}^{(\nu, \mu)}]_p|^2 ]$. It follows
immediately that $\hat{\boldsymbol{R}}_ {\boldsymbol{Y}
\boldsymbol{Y}} \doteq N_r P \sigma _x^2 \boldsymbol{B}
(\tilde{\varepsilon}) \boldsymbol{B}^H (\tilde{\varepsilon})$. By
invoking the definition of $[\boldsymbol{c}]_q$, we can further
obtain
\begin{equation}
[\boldsymbol{c}]_q \doteq N_r P \sigma _x^2 (Q-q) \tilde{z}^{-q}
\sum\limits_{\mu =0} ^{N_t-1} { z_{\mu}^{-q} }, \ 1 \leq q \leq Q-1,
\end{equation}
where $\tilde{z} = e^{j2\pi \tilde{\varepsilon} /Q}$. By
substituting the above result into (\ref{ghz}), the polynomial
coefficient corresponding to $z^{-q}$ at both sides of (\ref{ghz})
can be calculated to be
\begin{equation}\label{coeff}
N_r P \sigma _x^2 q(Q-q)\tilde{z}^{q} \sum\limits_{\mu=0} ^{N_t-1}
\sum\limits_{\mu '=0} ^{N_t-1} \left( {z_{\mu}}/{z_{\mu '}} \right)
^q ,
\end{equation}
where we have utilized the following property $z_{\mu} ^Q = 1$. This
completes the proof.

\section{} 

    This appendix presents the proof of (\ref{gzl0}). With CBTS
and (\ref{coeff}), we can readily obtain
\begin{multline}
\boldsymbol{c}^T \left\{ \left[\sum\limits _{\mu=0} ^{N_t-1}
{\boldsymbol{b} (z_{\mu})} \right] \odot \boldsymbol{b}(\tilde{z})
\odot \boldsymbol{q} \right\} \\
= N_r P \sigma _x^2 (\boldsymbol{q}
\odot \boldsymbol{ \Psi} )^H [(Q-\boldsymbol{q}) \odot \boldsymbol{
\Psi} ],
\end{multline}
where $\boldsymbol{ \Psi} = [\Psi(0), \Psi(1), \cdots, \Psi(q),
\cdots, \Psi(Q-1)]^T$, $\Psi(q) = e^{-j 2\pi q i_0 /Q}
\sum\nolimits_{\mu =0}^ {N_t -1} { e^{j 2 \pi q i_{\mu} /Q} }$. From
the definition of $\boldsymbol{\Psi}$, we have
\begin{equation}
\boldsymbol{\Psi} = \boldsymbol{\Phi} \boldsymbol{1}_{N_t},
\end{equation}
where $\boldsymbol{\Phi} = [\boldsymbol{\phi}_0,
\boldsymbol{\phi}_1, \cdots, \boldsymbol{\phi}_{N_t-1}]$,
$\boldsymbol{\phi}_{\mu} = [1, e^{j 2 \pi (i_{\mu}-i_0) /Q}, \cdots,
e^{j 2\pi q (i_{\mu}-i_0) /Q}, \cdots, e^{j 2 \pi (Q-1)
(i_{\mu}-i_0) /Q}]^T$. Since $\boldsymbol{\Phi}$ is a Vandermonde
matrix, it is of full-rank (rank $N_t$) with $N_t < Q$.
Consequently, $\boldsymbol{ \Psi}$ cannot be the all zero vector and
then $\boldsymbol{c}^T \left\{ \left[\sum\nolimits _{\mu=0} ^{N_t-1}
{\boldsymbol{b} (z_{\mu})} \right] \odot \boldsymbol{b}(\tilde{z})
\odot \boldsymbol{q} \right\} > 0$. From (\ref{equ6}), it follows
immediately that $f'(\tilde{z}) = 0$. This completes the proof.

\bibliographystyle{IEEEtran}

\begin{thebibliography}{10}
\providecommand{\url}[1]{#1} \csname url@rmstyle\endcsname
\providecommand{\newblock}{\relax} \providecommand{\bibinfo}[2]{#2}
\providecommand\BIBentrySTDinterwordspacing{\spaceskip=0pt\relax}
\providecommand\BIBentryALTinterwordstretchfactor{4}
\providecommand\BIBentryALTinterwordspacing{\spaceskip=\fontdimen2\font
plus \BIBentryALTinterwordstretchfactor\fontdimen3\font minus
  \fontdimen4\font\relax}
\providecommand\BIBforeignlanguage[2]{{%
\expandafter\ifx\csname l@#1\endcsname\relax
\typeout{** WARNING: IEEEtran.bst: No hyphenation pattern has been}%
\typeout{** loaded for the language `#1'. Using the pattern for}%
\typeout{** the default language instead.}%
\else \language=\csname l@#1\endcsname \fi #2}}

\bibitem{Moose}
P.~Moose, ``A technique for orthogonal frequency division
multiplexing
  frequency offset correction,'' \emph{IEEE Trans. Commun.}, vol.~42, pp.
  2908--2914, Oct. 1994.

\bibitem{David01}
D.~Huang and K.~B. Letaief, ``Carrier frequency offset estimation
for {OFDM}
  systems using null subcarriers,'' \emph{IEEE Trans. Commun.}, vol.~54, no.~5,
  pp. 813--823, May 2006.

\bibitem{FeifeiGao}
F.~Gao and A.~Nallanathan, ``Blind maximum likelihood {CFO}
estimation for
  {OFDM} systems via polynomial rooting,'' \emph{IEEE Signal Processing Lett.},
  vol.~13, no.~2, pp. 73--76, Feb. 2006.

\bibitem{Xiaoli}
X.~Ma, M.~K. Oh, G.~B. Giannakis, and D.~J. Park, ``Hopping pilots
for
  estimation of frequency-offset and multi-antenna channels in {MIMO OFDM},''
  \emph{IEEE Trans. Commun.}, vol.~53, no.~1, pp. 162--172, Jan. 2005.

\bibitem{Simoens}
F.~Simoens and M.~Moeneclaey, ``Reduced complexity data-aided and
code-aided
  frequency offset estimation for flat-fading {MIMO} channels,'' \emph{IEEE
  Trans. Wireless Commun.}, vol.~5, no.~6, pp. 1558--1567, June 2006.

\bibitem{Yxjiang3}
Y.~X. Jiang, X.~Q. Gao, X.~H. You, and W.~Heng, ``Training sequence
assisted
  frequency offset estimation for {MIMO OFDM},'' in \emph{Proc. IEEE ICC'06},
  vol.~12, Istanbul, Turkey, June 2006, pp. 5371--5376.

\bibitem{YxjiangTWC}
Y.~X. Jiang, H.~Minn, X.~Q. Gao, X.~H. You, and Y.~Li, ``Frequency
offset
  estimation and training sequence design for {MIMO OFDM},'' \emph{accepted, to
  be published in IEEE Trans. Wireless Commun.}, Jan. 2008.

\bibitem{Press}
W.~H. Press, \emph{Numerical Recipes in C++: the Art of Scientific
  Computing}.\hskip 1em plus 0.5em minus 0.4em\relax Cambridge: Cambridge
  University Press, 2002.

\bibitem{Chu}
D.~Chu, ``Polyphase codes with good periodic correlation
properties,''
  \emph{IEEE Trans. Inform. Theory}, vol.~18, pp. 531--532, July 1972.

\bibitem{Stuber}
G.~L. Stuber, J.~R. Barry, S.~Mclaughlin, Y.~Li, M.~A. Ingram, and
T.~G. Pratt,
  ``Broadband {MIMO-OFDM} wireless communications,'' \emph{Proceedings of the
  IEEE}, vol.~92, no.~2, pp. 271--294, Feb. 2004.

\bibitem{Gini}
F.~Gini and R.~Reggiannini, ``On the use of {Cramer-Rao-like} bounds
in the
  presence of random nuisance parameters,'' \emph{IEEE Trans. Commun.},
  vol.~48, no.~12, pp. 2120--2126, Dec. 2000.

\bibitem{Minn_Periodic}
H.~Minn, X.~Fu, and V.~K. Bhargava, ``Optimal periodic training
signal for
  frequency offset estimation in frequency-selective fading channels,''
  \emph{IEEE Trans. Commun.}, vol.~54, no.~6, pp. 1081--1096, June 2006.

\end{thebibliography}

\end{document}